\documentclass[twocolumn]{article} % twocolumn

\usepackage{graphicx}
\usepackage{graphbox}
\usepackage{cite}
\usepackage{amssymb}
\usepackage{latexsym}
\usepackage{graphicx}
\usepackage{subfigure}
\usepackage{amsmath}
\usepackage{mathtools}
\usepackage{url}
\usepackage{pgf,tikz}
\usepackage{tikz}
\usetikzlibrary{spy}
\usepackage{xcolor}
\usepackage{algorithm}%
\usepackage{algorithmicx}%
\usepackage{algpseudocode}%
\usepackage{listings}%
\usepackage{graphbox}
\usepackage{tabularx}
\newcolumntype{L}{>{\raggedright\arraybackslash}X}
\graphicspath{ {images/} }

\begin{document}
\sloppy

\title{Learning-Based and Quality Preserving Super-Resolution of Noisy Images}%

\makeatletter
%\begin{center}
\onecolumn
{\fontsize{18pt}{20pt}\selectfont\bfseries\@title\par}
Simone Cammarasana
 \footnote{
\textbf{Simone Cammarasana}
CNR-IMATI, Via De Marini 6, Genova, Italy \\
simone.cammarasana@ge.imati.cnr.it
}
  , 
 Giuseppe Patan\`e
  \footnote{
 \textbf{Giuseppe Patan\`e} 
 CNR-IMATI, Via De Marini 6, Genova, Italy 
}

%\end{center}
\makeatother

\begin{abstract}
Several applications require the super-resolution of noisy images and the preservation of geometrical and texture features. State-of-the-art super-resolution methods do not account for noise and generally enhance the output image's artefacts (e.g., aliasing, blurring). 
We propose a learning-based method that accounts for the presence of noise and preserves the properties of the input image, as measured by quantitative metrics (e.g., normalised crossed correlation, normalised mean squared error, peak-signal-to-noise-ration, structural similarity feature-based similarity, universal image quality). We train our network to up-sample a low-resolution noisy image while preserving its properties. We perform our tests on the Cineca Marconi100 cluster, at the 26th position in the ``\emph{top500}'' list.
The experimental results show that our method outperforms learning-based methods, has comparable results with standard methods, preserves the properties of the input image as contours, brightness, and textures, and reduces the artefacts. As average quantitative metrics, our method has a PSNR value of 23.81 on the super-resolution of Gaussian noise images with a 2X up-sampling factor. In contrast, previous work has a PSNR value of 23.09 (standard method) and 21.78 (learning-based method).
Our learning-based and quality-preserving super-resolution improves the high-resolution prediction of noisy images with respect to state-of-the-art methods with different noise types and up-sampling factors.

\end{abstract}

\section{Introduction}
The super-resolution of 2D images is primarily studied and widespread in many applications, such as biomedicine~\cite{lapini2014comparison}, astronomy~\cite{puschmann2005super}, and industrial~\cite{qin2022progressive} context. In the literature, several methods guarantee excellent results in terms of reconstruction accuracy~\cite{singh2020survey}. However, most of the current super-resolution methods do not account for noise in the image. At the same time, the preservation of the features and visual quality of the input data is affected by an underlying noise distribution (Sect.~\ref{SEC:RELATEDWORK}). In contrast, these super-resolution methods generally smooth the noise with a blurring effect~\cite{villar2021deep},~\cite{mushtaq2022super} or apply a denoising filter before the super-resolution~\cite{han2021Deep}. 

On the one hand, the denoise effect may be required in image processing; on the other hand, noise reduction alters the visual quality of the super-resolution image. For example, ultrasound images are affected by speckle noise, which is generated by the superimposition of ultrasound waves. The super-resolution of ultrasound images improves anatomical structures' visibility, keeps the quality of the input data unchanged, makes the high-resolution image visually close to the corresponding low-resolution, and avoids blurring artefacts, allowing the physician better to visualise anatomical features in the image~\cite{cammarasana2023super}. Furthermore, ultrasound images are eventually processed with specialised denoising filters applied after the super-resolution to remove the noise component while preserving the anatomical features~\cite{PMID:35672630}. It is relevant to generate high-resolution images that keep the visual quality regarding noise properties, main geometries and grey-scale values and eventually apply the dedicated denoise filter afterwards. 

\begin{figure*}[t]
%\captionsetup{width=0.95\columnwidth}
\centering
\begin{tabular}{cc|cc}
\includegraphics[width=.22\textwidth]{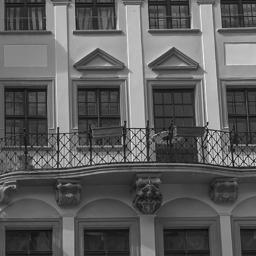} &
\includegraphics[width=.22\textwidth]{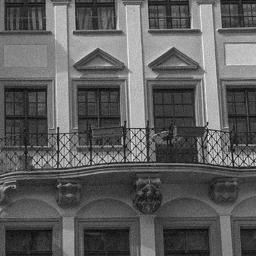} &
\includegraphics[width=.11\textwidth]{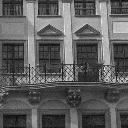} &
\includegraphics[width=.22\textwidth]{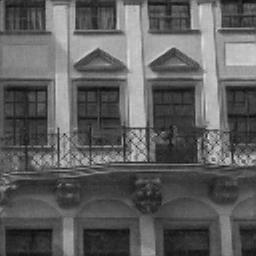} \\
Ground-truth & Noisy & Low-resolution & Super-resolution 2X
\end{tabular}
\caption{Given a noisy low-resolution image (e.g.,~$128 \times 128$, Gaussian noise), we generate a high-resolution (e.g.,~$256 \times 256$) prediction that matches the noisy high-resolution input image.\label{FIG:TEASER}}
\end{figure*}

In this context, we aim to generate a high-resolution image from a low-resolution noisy image, preserving the visual quality and the quantitative similarity of the input data without altering the noise distribution. To this end (Fig.~\ref{FIG:TEASER}), we define a novel super-resolution of noisy images based on a learning model that accounts for the presence of the noise and preserves the similarity of the high-resolution image in terms of visual quality and quantitative metrics (e.g., normalised crossed correlation, normalised mean squared error, peak-signal-to-noise-ration, structural similarity feature-based similarity, universal image quality). In contrast, state-of-the-art super-resolution methods do not account for the noise distribution to preserve the similarity with the high-resolution image. For the learning of our model, we define a data set of ground-truth, noisy, and noisy down-sampled synthetic 2D images, train our network to up-sample the low-resolution noisy image, and match the high-resolution noisy image. At the same time, the prediction has to preserve the noise properties with respect to the ground-truth image. We specialise our networks to different up-sampling factors, i.e., 2X and 4X, and noise, i.e., speckle and Gaussian (Sect.~\ref{SEC:METHOD}). Our approach is general to specialise with additional noise types (e.g., Poisson) and up-sampling factors (e.g., 8X). 

As the main result, our method improves the super-resolution of noisy images with respect to state-of-the-art standard and learning-based methods. We evaluate several quantitative measures for local, structural, and quality similarity. For example, we achieve a PSNR average value of 20.64 with 4X up-sampling factor and Gaussian noise, while previous works achieve an average 20.11 value; also, we gain an average mean-squares error value of 247 with 2X up-sampling factor and speckle noise, while state-of-the-art methods achieve more than 400 as an average value. We evaluate the preservation of image properties, such as brightness, contours, and textures. Our super-resolution has comparable results with the standard method and improves learning-based methods on ultrasound images affected by speckle noise. We also analyse the distribution of the generated noise with respect to the ground-truth image; our method generally better preserves the noise distribution. Finally, we discuss the conclusion and future works (Sect.~\ref{SEC:CONCLUSION}). Trained networks for 2X and 4X up-sampling factors, Gaussian and speckle noise, are available at \url{https://github.com/cammarasana123/noise-SuperResolution}.

\section{Related work\label{SEC:RELATEDWORK}}
We discuss deep learning and standard methods for the super-resolution of 2D images.
\paragraph*{Learning-based super-resolution}
In the last years, deep learning methods for super-resolution are widespread. The \emph{statistical prediction model} (SPM) applies a sparse representation of patch pairs over two dictionaries, one for both low-resolution (LR) and high-resolution (HR) images~\cite{peleg2014statistical}, and captures the statistical dependencies between the sparsity patterns of the low and high-resolution coefficients of the corresponding representations. In~\cite{dong2014learning,dong2015image}, a fully convolutional neural network exploits a large filter size in the non-linear mapping layer. It applies three colour channels to train and predict HR images from LR images. A data clustering groups patches, and a Bayes strategy selects patches and provides a fast super-resolution based on external learning~\cite{salvador2015naive}. The main novelties of the \emph{enhanced deep super-resolution} (EDSR) network~\cite{lim2017enhanced} are a simplification of the conventional residual network architectures and a multi-scale super-resolution network that reduces the model size.

A \emph{super-resolution generative adversarial network} (SRGAN)~\cite{ledig2017photo} applies a deep residual network with skip-connection and a perceptual loss between generated and target images. The reduction of artefacts of the previous method is addressed by the Enhanced SRGAN~\cite{10.1007/978-3-030-11021-5_5}, which improves the network architecture, the adversarial and the perceptual loss, removes the batch normalisation layer, and applies the residual scaling and smaller initialisation values. The perceptual quality of ESRGAN is improved by the ESRGAN+ method~\cite{rakotonirina2020esrgan+} through a novel \emph{Residual-in-Residual Dense Residual} block, which increases the network capacity without affecting its complexity. The introduction of weight normalisation and wider features before rectified linear unit activation function~\cite{yu2020wide} achieves good super-resolution results with a low computational cost. Additional methods, classified according to supervised/unsupervised approach and domain-specific applications, are discussed in~\cite{wang2020deep}.
\begin{figure}[t]
%\captionsetup{width=0.95\columnwidth}
\centering
\includegraphics[width=.94\columnwidth]{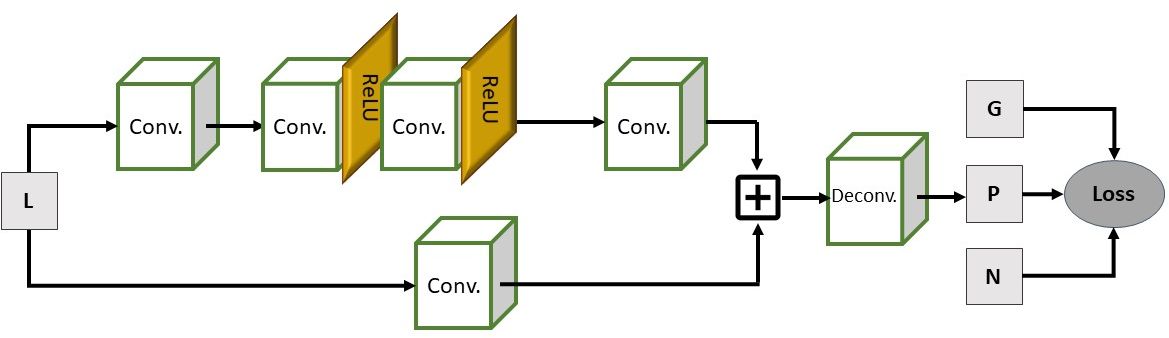}  \\
\caption{Network architecture: noisy low-resolution ($\mathbf{L}$), ground-truth ($\mathbf{G}$), noisy high-resolution ($\mathbf{N}$), prediction ($\mathbf{P}$). \label{FIG:NETWORK}}
\end{figure}
\paragraph*{Standard super-resolution}
We refer to non-learning-based methods as standard methods.
The interpolating super-resolution with cubic kernels (i.e., \emph{cubic convolution}~\cite{keys1981cubic}, CC in short) offers high accuracy with a low computational cost through appropriate boundary conditions and constraints on the kernel functions. The interpolated values are computed as the weighted average of pixels in the~$2\times2$ (bilinear,~\cite{gribbon2003real}) or~$4\times4$ (bi-cubic,~\cite{mahale2014hardware}) neighbourhood. A fast implementation of bilinear and bi-cubic interpolations~\cite{khaledyan2020low} applies to \emph{Field Programmable Gate Array} (FPGA), reducing the computational complexity and the FPGA resources while providing an excellent trade-off between image quality and calculation simplicity. After the training of two dictionaries for LR and HR patches~\cite{yang2010image}, the similarity of the sparse representation of LR and HR patches with the respective dictionary is exploited to generate the high-resolution image. Anchored neighbourhood regression~\cite{timofte2013anchored} and its improved version~\cite{timofte2014a+} learn a regression to correlate LR and HR images for each atom of the dictionary and precomputed neighbourhood. The search of recursive patches within an image~\cite{huang2015single} is extended by allowing geometric variations, incorporating the geometry by localising planes and accounting plane parameters to estimate the deformation of recurring patches. Bivariate rational fractal interpolation~\cite{zhang2018single} improves the approximation results with respect to polynomial kernels, preserving image edges and textures.
\begin{figure*}[t]
%\captionsetup{width=0.95\columnwidth}
\centering
\begin{tabular}{ccc}
\includegraphics[width=.27\textwidth]{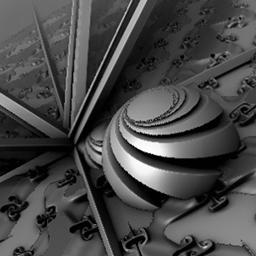} &
\includegraphics[width=.27\textwidth]{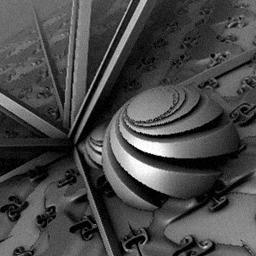} &
\includegraphics[width=.27\textwidth]{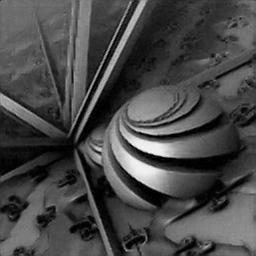} \\
Ground-truth & Noisy & Ours \\
\includegraphics[width=.27\textwidth]{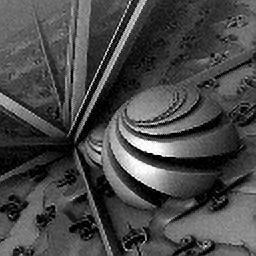} &
\includegraphics[width=.27\textwidth]{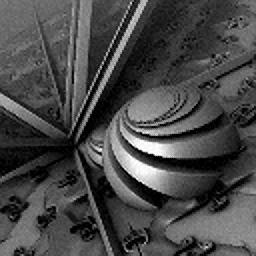} &
\includegraphics[width=.27\textwidth]{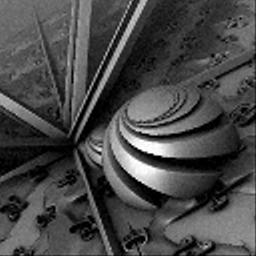}  \\
EDSR & SPM & CC
\end{tabular}
\caption{Comparison among super-resolution methods, 2X up-sampling factor, Gaussian noise.\label{FIG:2XTEST}}
\end{figure*}
\section{Learning-based and quality-preserving image super-resolution\label{SEC:METHOD}}
We describe our learning-based super-resolution method for 2D noisy images (Sect.~\ref{SEC:MODEL}), the data set and quantitative metrics for the super-resolution network (Sect.~\ref{SEC:METRICS}), and the experimental results (Sect.~\ref{SEC:RESULTS}).
\begin{figure*}[t]
%\captionsetup{width=0.95\columnwidth}
\centering
\begin{tabular}{ccc}
\includegraphics[width=.27\textwidth]{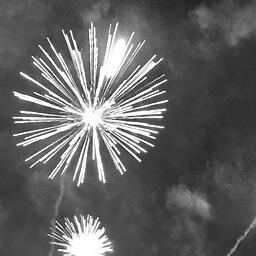} &
\includegraphics[width=.27\textwidth]{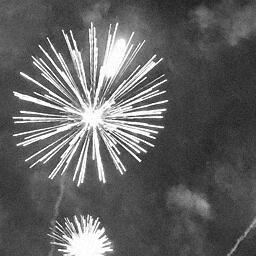} &
\includegraphics[width=.27\textwidth]{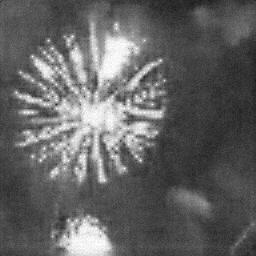} \\
Ground-truth & Noisy & Ours \\
\includegraphics[width=.27\textwidth]{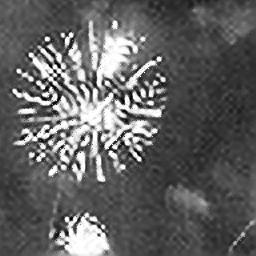} &
\includegraphics[width=.27\textwidth]{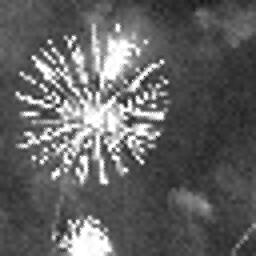} &
\includegraphics[width=.27\textwidth]{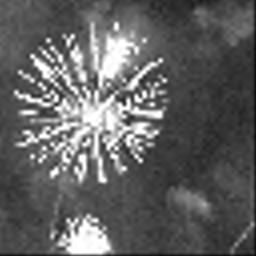}  \\
EDSR & SPM & CC
\end{tabular}
\caption{Comparison among super-resolution methods, 4X up-sampling factor, Gaussian noise.\label{FIG:4XTEST}}
\end{figure*}
\subsection{Learning model\label{SEC:MODEL}}
We select WDSR~\cite{yu2020wide}, an architecture that exploits residual blocks since it improves the prediction of images where the difference between the input and the target is small. We propose a customised version of this network: \emph{custom-WDSR}. In particular, our network architecture is a variant of WDSR-A, where the expansion of the features before the rectified linear unit (ReLU) activation allows more information to pass through while preserving the non-linearity of the network. After normalising the data, we apply a 2D convolution and a weighted normalisation that improves the conditioning of the optimisation problem and, thus, the convergence. Then, we apply eight residual blocks with wide activation, where each comprises two convolution layers with ReLU activation and a final 2D convolution with a weighted normalisation layer. Finally, we combine residual blocks and convolution layers, apply deconvolution to interpolate the missing lines and columns, and denormalisation to match the target image. The kernel filter size depends on the up-sampling factor:~$(3 \times 3)$ in 2X up-sampling, and~$(5 \times 5)$ in 4X up-sampling. With this setting, the total number of trained parameters is 889K for 2X network and 253K for 4X network (Fig.~\ref{FIG:NETWORK}).

The network input is a noisy, low-resolution image, while the target is the corresponding high-resolution image. The loss function accounts for both the loss between the predicted image and the target image (i.e., the noisy image) and the log-likelihood of the generated noise with respect to the input noise. In particular, we define the prediction ($\mathbf{P}$), noisy ($\mathbf{N}$) and ground-truth ($\mathbf{G}$) images, and the loss as
\begin{equation}
\label{EQ:LOSS}
\mathcal{L} = \| \mathbf{P} - \mathbf{N} \|_F + \lambda \frac{1}{m} \sum{\log(L(\mathbf{P}-\mathbf{G},\Theta))},
\end{equation}
where~$\Theta$ is the known noise distribution (e.g., Gaussian) and~$m$ is the number of pixels of the high-resolution image. The first term (i.e.,~$\| \mathbf{P}-\mathbf{N}\|_F$) trains the network to match the target high-resolution noisy image, while the second term trains the network to generate a prediction whose noise distribution with respect to the ground-truth image matches the~$\Theta$ distribution; the second term is computed with the \emph{log-loss} function~$L$, and it is weighted through the parameter~$\lambda=-10$. The~$\lambda$ term is negative as we maximise the noise distribution properties with respect to the~$\Theta$ distribution.
 As additional parameters for the training, we apply the Adam optimiser with a learning rate of 0.001, a maximum number of epochs of 60, and an early stopping criterion by monitoring the validation loss. We train our networks with 400 images for the training data set, 70 for the validation data set, and 30 for the test data set.
\begin{figure}[t]
%\captionsetup{width=0.95\columnwidth}
\centering
\begin{tabular}{cc}
\includegraphics[width=.45\columnwidth]{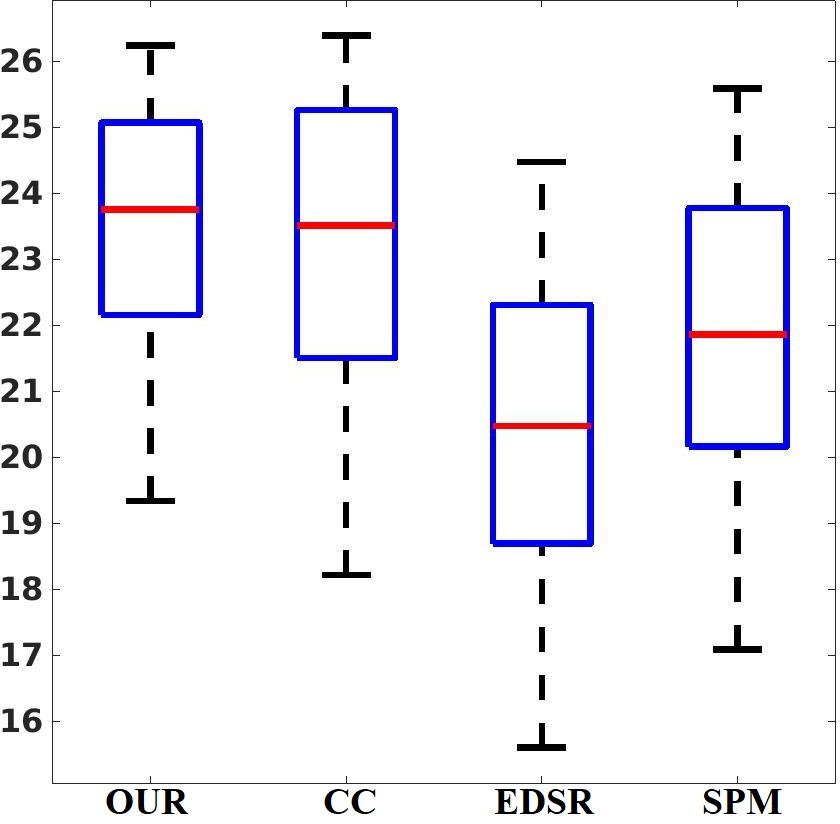} &
\includegraphics[width=.45\columnwidth]{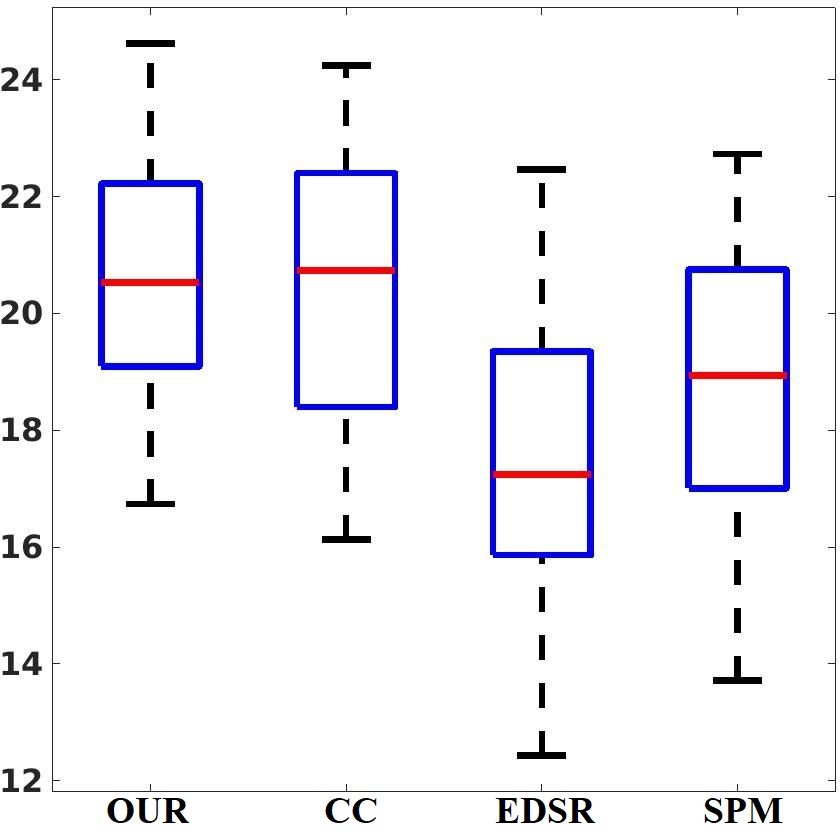} \\
2X & 4X
\end{tabular}
\caption{2X and 4X box plot of PSNR with Gaussian noise. \label{FIG:BOXPLOTGAUSS}}
\end{figure}
\subsection{Data sets and metrics\label{SEC:METRICS}}
We account for a data set of synthetic images composed of 500 images taken from the \emph{Imagenet} data set~\cite{russakovsky2015imagenet}. Given the ground-truth image~$\mathbf{G}$, we apply the artificial noise (e.g., Gaussian, speckle) generating the noisy image~$\mathbf{N}$. Then, we down-sample the~$\mathbf{N}$ image to the noisy low-resolution~$\mathbf{L}$ image with a down-sampling factor. We apply two different down-sampling factors of~$k=2$, where we remove one row every two and one column every two, and~$k=4$, where we remove three columns and three rows every four. We generate separated data sets in terms of up-sampling factor and noise type, as the specialisation of the trained networks improves the accuracy of the reconstruction of the target image.
\begin{table}[t]
%\captionsetup{width=0.95\columnwidth}
\centering
\caption{Concerning the 2X up-sampling factor results on Gaussian noise images, we report the metrics computed between target and super-resolution methods as average values on the test data set. The best results are in bold.\label{TAB:2XTEST}}
\begin{tabular}{c|cccc}
Methods & CC & EDSR & SPM & OUR \\ \hline
MSE & 406.76 & 809.74 & 683.21  &~$\mathbf{304.11}$\\
NRMSE &0.147 & 0.229 & 0.195 &~$\mathbf{0.116}$ \\
NCC & 0.876 & 0.825& 0.848&~$\mathbf{0.891}$ \\
PSNR &23.09 & 20.47 &21.78 &$\mathbf{23.81}$ \\
SSIM &0.823 &0.718 & 0.773 &$\mathbf{0.831}$ \\
FSIM & 0.914 & 0.860 & 0.886 &$\mathbf{0.915}$ \\
UIQ &~$\mathbf{0.998}$ & 0.984 & 0.979 & 0.997 
\end{tabular}
\end{table}
\paragraph{Quality metrics}
The quality preservation of our super-resolution is measured through quantitative metrics.
The predicted image~$\mathbf{P}$ is compared with~$\mathbf{N}$ to measure the super-resolution accuracy and with~$\mathbf{G}$ to measure the generated noise properties. Our comparison accounts for the visual similarity of the super-resolution images (predicted and noisy) to quantify the preservation of the geometries and features and evaluate several metrics for local, structural, and texture similarity. Given the noisy and the predicted image on~$m$ points, we compute the
\begin{itemize}
\item \emph{normalised crossed correlation} \mbox{$\textrm{NCC} = \frac{ \sum_{i=1}^{m} [\mathbf{N}(i) - \overline{\mathbf{N}}][\mathbf{P}(i) - \overline{\mathbf{P}}]}{ \left [ \sum_{i=1}^{m} [\overline{\mathbf{N}} - \overline{\mathbf{N}}]^{2} \right ]^{1/2} \left [ \sum_{i=1}^{m}[\mathbf{P}_i - \overline{\mathbf{P}}]^{2} \right ]^{1/2}}$}, where~$\overline{\mathbf{N}}$ and~$\overline{\mathbf{P}}$ are the average values of the two images;
\item the \emph{mean squared error} \mbox{$\textrm{MSE} = \frac{1}{m}\sum_{i=1}^{m}[\mathbf{N}(i) - \mathbf{P}(i)]^2$} and the \emph{normalised mean squared error} \mbox{$\textrm{NRMSE} = \left [\frac{\sum_{i=1}^{m}[\mathbf{N}(i) - \mathbf{P}(i)]^{2}} {\sum_{i=1}^{m}[\mathbf{N}(i)]^{2}}\right ]^{1/2}$};
\item the \emph{peak-signal-to-noise-ration} \mbox{$\textrm{PSNR}=10\log_{10}\frac{ (\max(\mathbf{N}))^2 }{MSE(\mathbf{N},\mathbf{P})}$};
\item the \emph{structural similarity} \mbox{$\textrm{SSIM}(\mathbf{P},\mathbf{N}) = l(\mathbf{P},\mathbf{N}) \times c(\mathbf{P},\mathbf{N}) \times s(\mathbf{P},\mathbf{N})$}, with \mbox{$l(\mathbf{P},\mathbf{N})=\frac{2\mu_{\mathbf{P}}\mu_{\mathbf{N}}+C_1}{\mu_{\mathbf{P}}^2 + \mu_{\mathbf{N}}^2 + C_1}$}, \mbox{$c(\mathbf{P},\mathbf{N}) = \frac{2\sigma_{\mathbf{P}}\sigma_{\mathbf{N}}+C_2}{\sigma_{\mathbf{P}}^2 + \sigma_{\mathbf{N}}^2 + C_2}$}, and \mbox{$s(\mathbf{P},\mathbf{N}) = \frac{\sigma_{\mathbf{PN}} + C_3}{ \sigma_{\mathbf{P}} \sigma_{\mathbf{N}} + C_3}$}, where~$\mu(\cdot)$ is the mean of~$(\cdot)$,~$\sigma(\cdot)$  is the standard deviation of~$(\cdot)$,~$\sigma_{\mathbf{PN}}$  is the covariance between~$\mathbf{P}$ and~$\mathbf{N}$, the positive constants~$C_1$,~$C_2$ and~$C_3$ are used to avoid a null denominator;
\item the \emph{feature-based similarity}~\cite{zhang2011fsim} \mbox{$\textrm{FSIM}=\frac{\sum_{i=1}^{m} S_L(i) \cdot PC(i)}{ \sum_{i=1}^{m} PC(i)}$}, where~$S_L = S_{PC} \cdot S_G$ is a combination of a similarity score of the phase congruency~$PC$~\cite{kovesi1999image} and the gradient magnitude~$G$;
\item the \emph{universal image quality} (UIQ)~\cite{wang2002universal} between~$\mathbf{N}$ and~$\mathbf{P}$ \mbox{$\textrm{UIQ}=\frac{4\sigma_{\mathbf{PQ}} \mu_\mathbf{P} \mu_\mathbf{Q}} {  (\sigma_\mathbf{P}^2 + \sigma_\mathbf{Q}^2) ( \mu_\mathbf{P}^2 + \mu_\mathbf{Q}^2 )}$}.
\end{itemize}
NCC, SSIM, and FSIM vary from 0 (worst case) to 1 (best case), PSNR varies from 0 (worst case) to~$+\infty$ (best case), MSE and NRMSE go from~$+\infty$ (worst case) to 0 (best case), UIQ varies from -1 (worst case) to +1 (best case). Finally, we perform a qualitative assessment of blurring, artefacts, and noise patterns, and analyse the histogram properties of the noisy component with respect to ground-truth image through~$(\mathbf{P} - \mathbf{G})$ and compare with the histogram properties of the input noise~$(\mathbf{N} - \mathbf{G})$.
\begin{figure*}[t]
%\captionsetup{width=0.95\columnwidth}
\centering
\begin{tabular}{cccc}
\hspace*{-0.375cm}
\includegraphics[width=.24\textwidth]{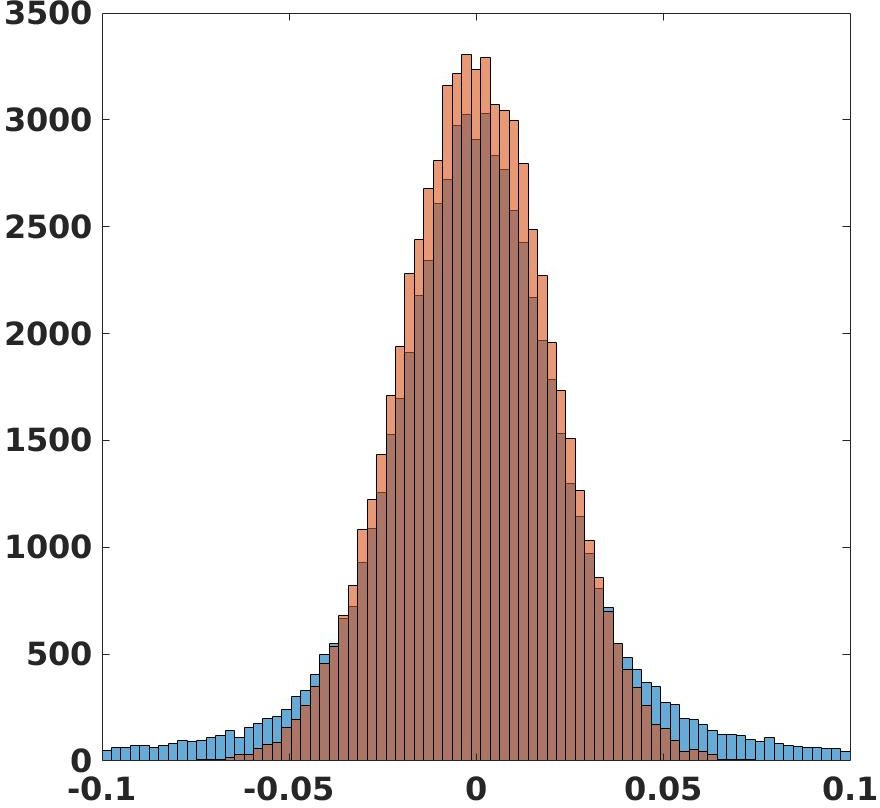}  &
\hspace*{-0.375cm}
\includegraphics[width=.24\textwidth]{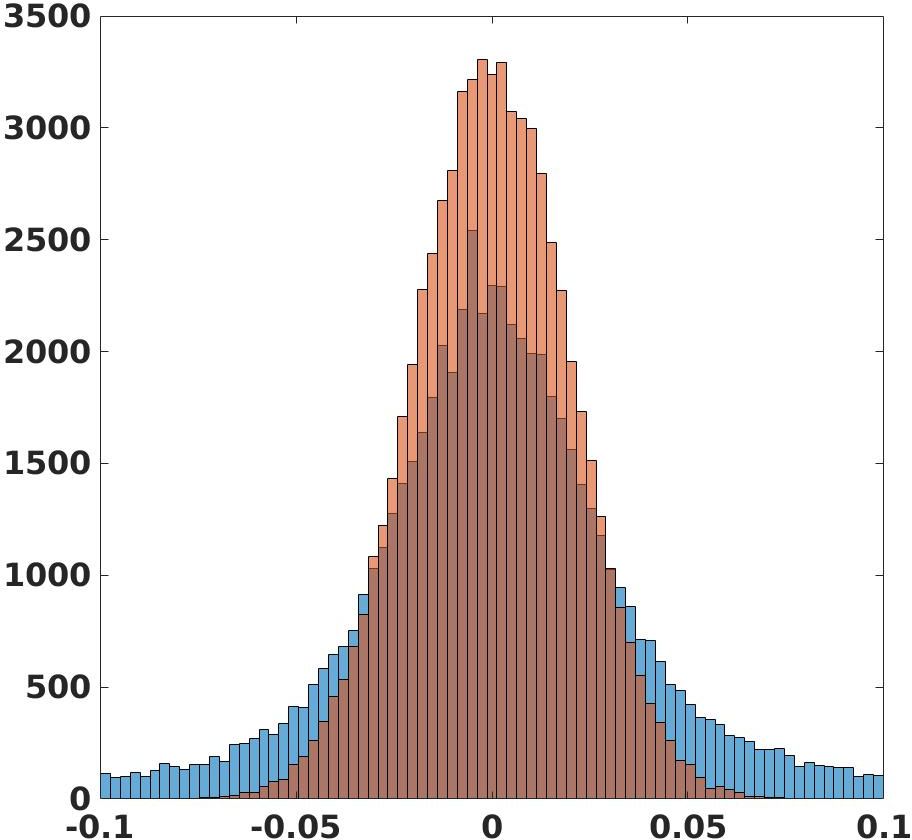} &
\hspace*{-0.375cm}
\includegraphics[width=.24\textwidth]{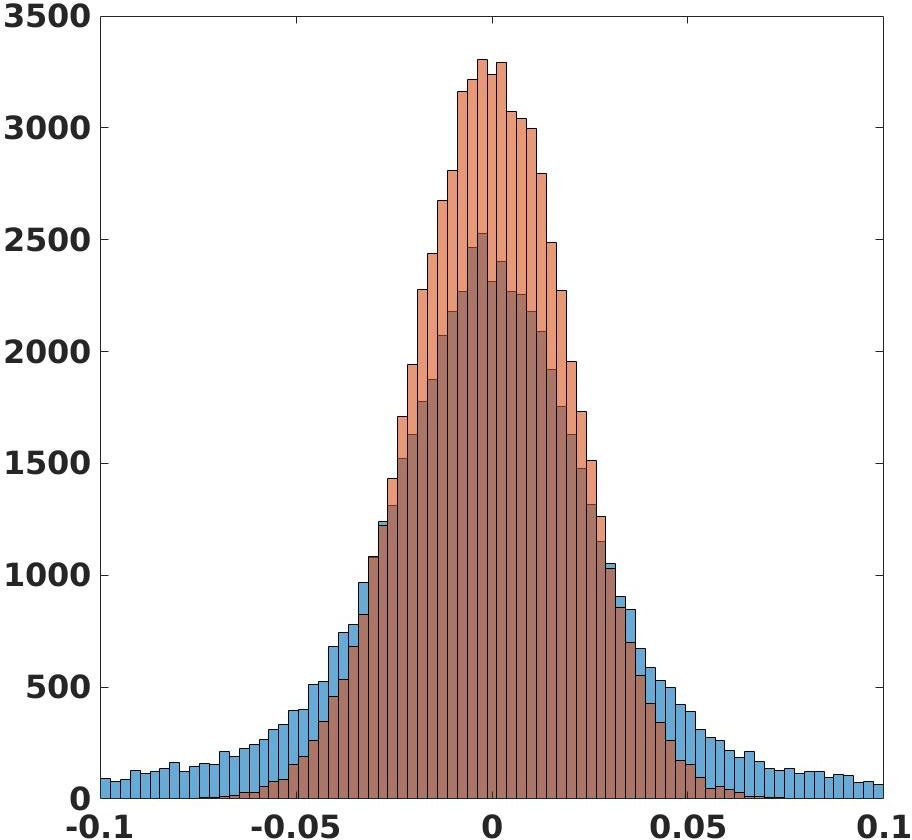} &
\hspace*{-0.375cm}
\includegraphics[width=.24\textwidth]{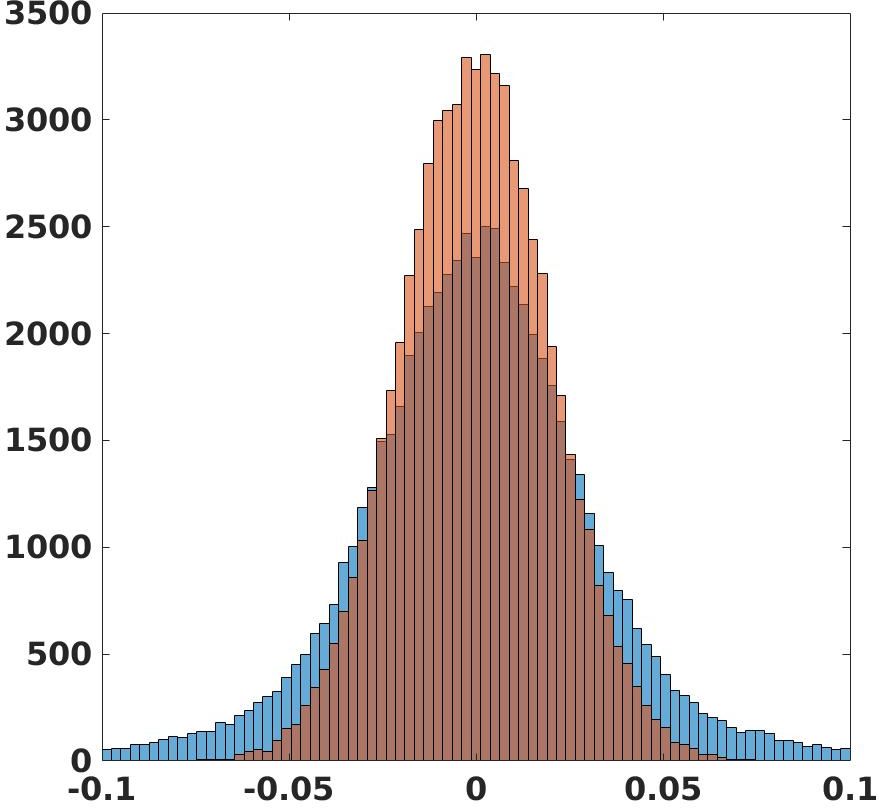}  \\
Our & EDSR & SPM & CC \\
\end{tabular}
\caption{2X up-sampling factor, Gaussian noise distribution: ground-truth (blue histogram) and super-resolution method (red histogram).\label{FIG:2XGAUSS}}
\end{figure*}
\subsection{Experimental results\label{SEC:RESULTS}}
We test our super-resolution method on synthetic images with 2X and 4X up-sampling factors and with Gaussian and speckle noise: the Gaussian noise has~$\mu=0$,~$\sigma=0.02$, and speckle noise has~$\sigma=0.02$. We compare our super-resolution with the methods CC~\cite{keys1981cubic}, EDSR~\cite{lim2017enhanced}, and SPM~\cite{peleg2014statistical}, described in Sect.~\ref{SEC:RELATEDWORK}.
\begin{figure*}[t]
%\captionsetup{width=0.95\columnwidth}
\centering
\begin{tabular}{cccc}
\hspace*{-0.375cm}
\includegraphics[width=.24\textwidth]{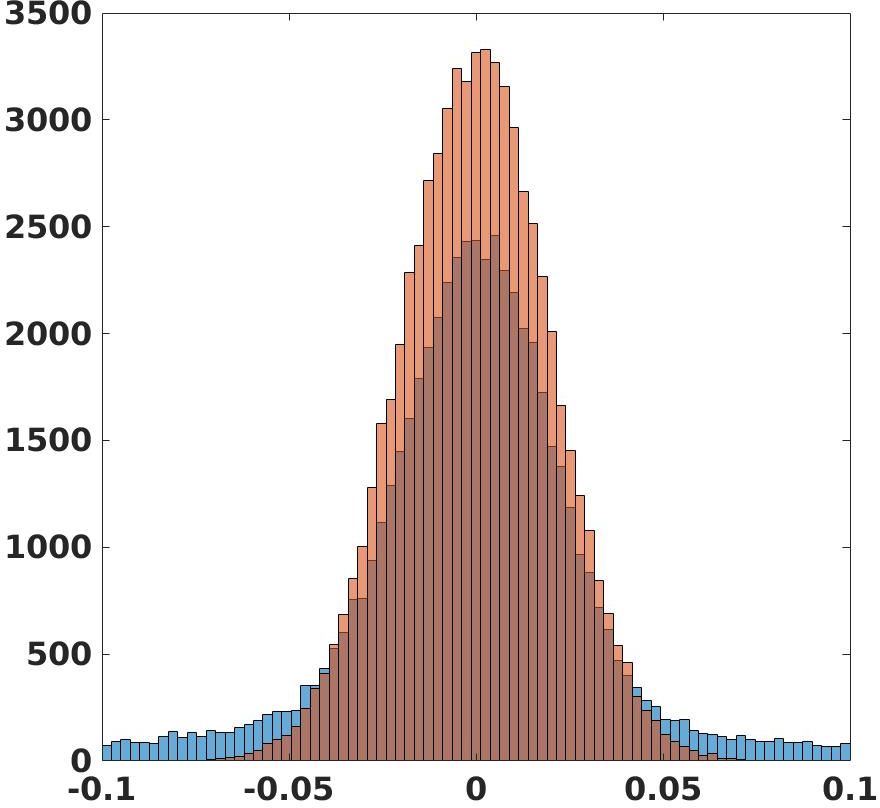} &
\hspace*{-0.375cm}
\includegraphics[width=.24\textwidth]{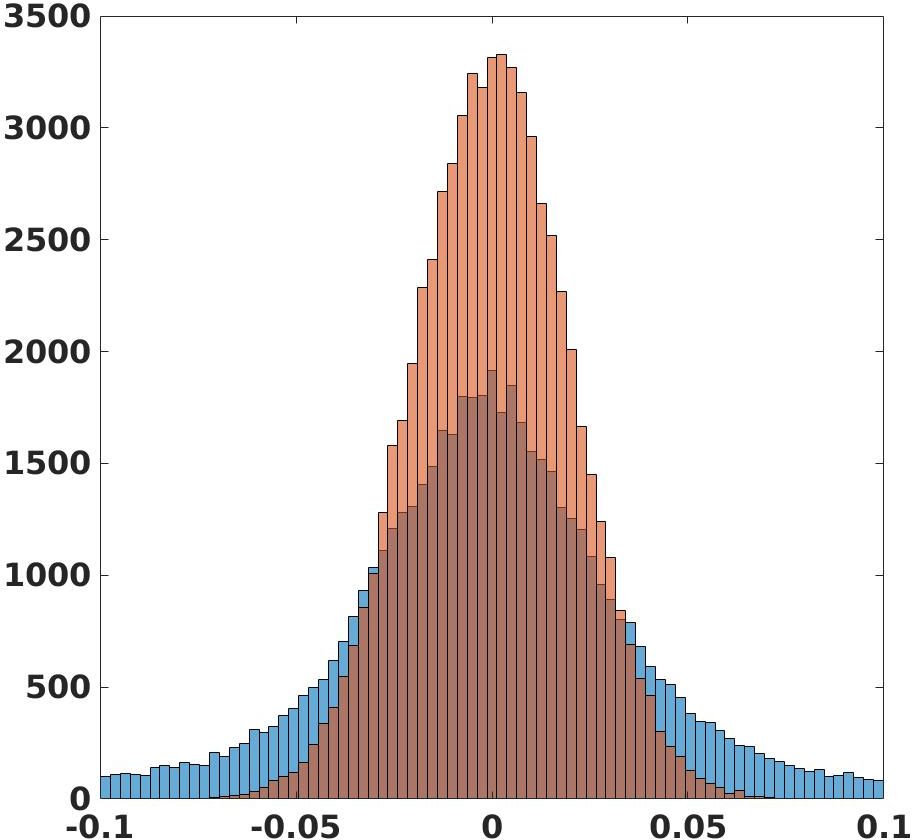} &
\hspace*{-0.375cm}
\includegraphics[width=.24\textwidth]{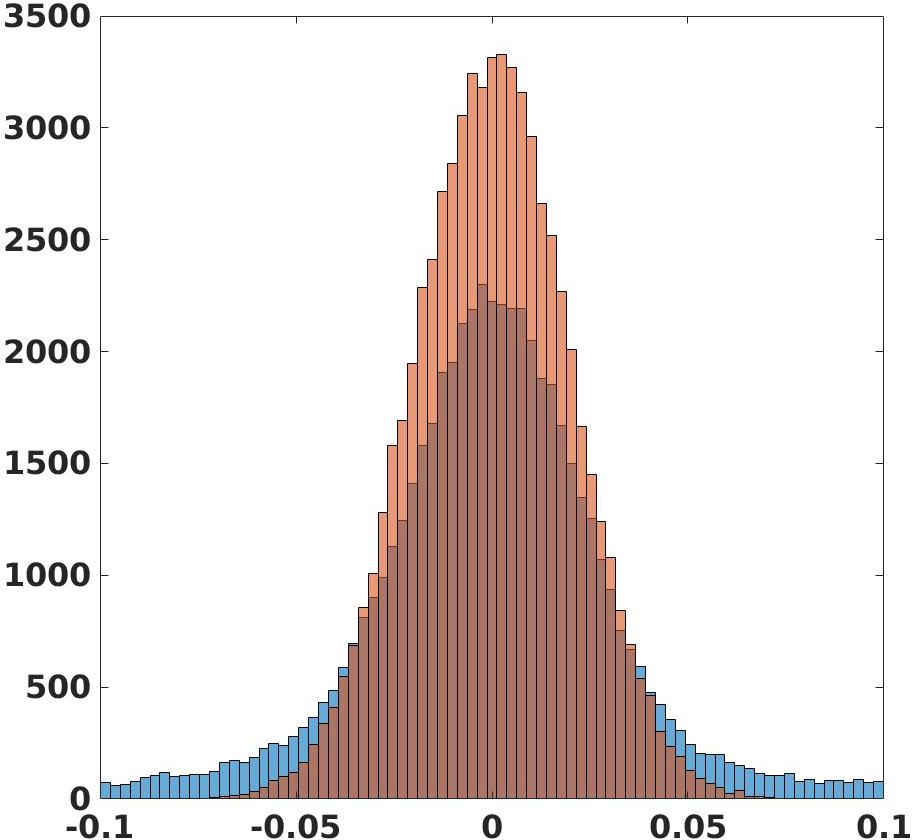} &
\hspace*{-0.375cm}
\includegraphics[width=.24\textwidth]{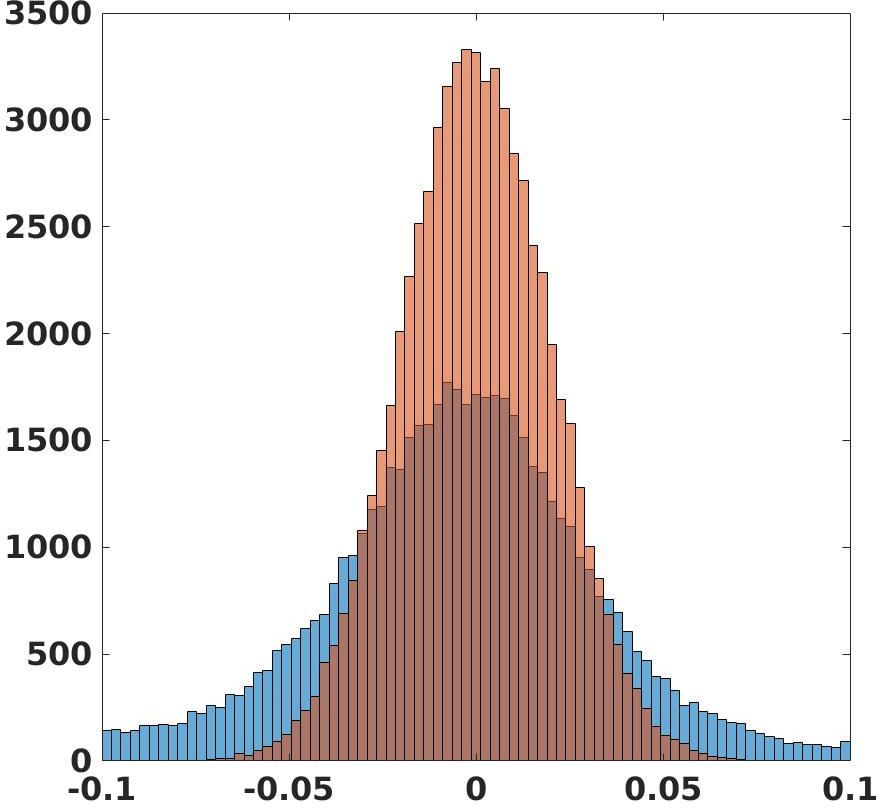}  \\
Our & EDSR & SPM & CC \\
\end{tabular}
\caption{4X up-sampling factor, Gaussian noise distribution: ground-truth (blue histogram) and super-resolution method (red histogram).\label{FIG:4XGAUSS}}
\end{figure*}
\paragraph{Gaussian noise}
Figs.~\ref{FIG:2XTEST},~\ref{FIG:4XTEST} show the super-resolution results with Gaussian noise at 2X and 4X, respectively. Our super-resolution better preserves the contours of the geometrical elements (e.g., the sphere's edges) and reduces the generation of artefacts, in particular the blurring effect on the sphere (c.f., CC in Fig.~\ref{FIG:2XTEST}) and the scattering effect of the clouds (c.f., EDSR in Fig.~\ref{FIG:4XTEST}). Our super-resolution also preserves texture, brightness, and saturation properties. None of the methods perfectly reconstruct the fireworks in the 4X example due to the large down-sampling factor of the input image.
\begin{table}[t]
%\captionsetup{width=0.95\columnwidth}
\centering
\caption{Concerning the 4X up-sampling factor results on Gaussian noise images, we report the metrics computed between target and super-resolution methods as the average value on the test data set. The best results are in bold.\label{TAB:4XTEST}}
\begin{tabular}{c|cccc}
Methods & CC & EDSR & SPM & OUR \\ \hline
MSE & 1112 & 1880 & 1372  &~$\mathbf{949}$\\
NRMSE &0.299 & 0.381 & 0.331 &~$\mathbf{0.277}$ \\
NCC & 0.822 & 0.721& 0.780&~$\mathbf{0.843}$ \\
PSNR &20.11 & 17.31 &18.74 &$\mathbf{20.64}$ \\
SSIM &~$\mathbf{0.591}$ &0.473& 0.553 &0.564 \\
FSIM &~$\mathbf{0.747}$ & 0.700 & 0.738 &0.739 \\
UIQ & 0.992 & 0.989 & 0.988 &~$\mathbf{0.994}$
\end{tabular}
\end{table}

\begin{figure*}[t]
%\captionsetup{width=0.95\columnwidth}
\centering
\begin{tabular}{cccc}
\hspace*{-0.375cm}
\includegraphics[width=.23\textwidth]{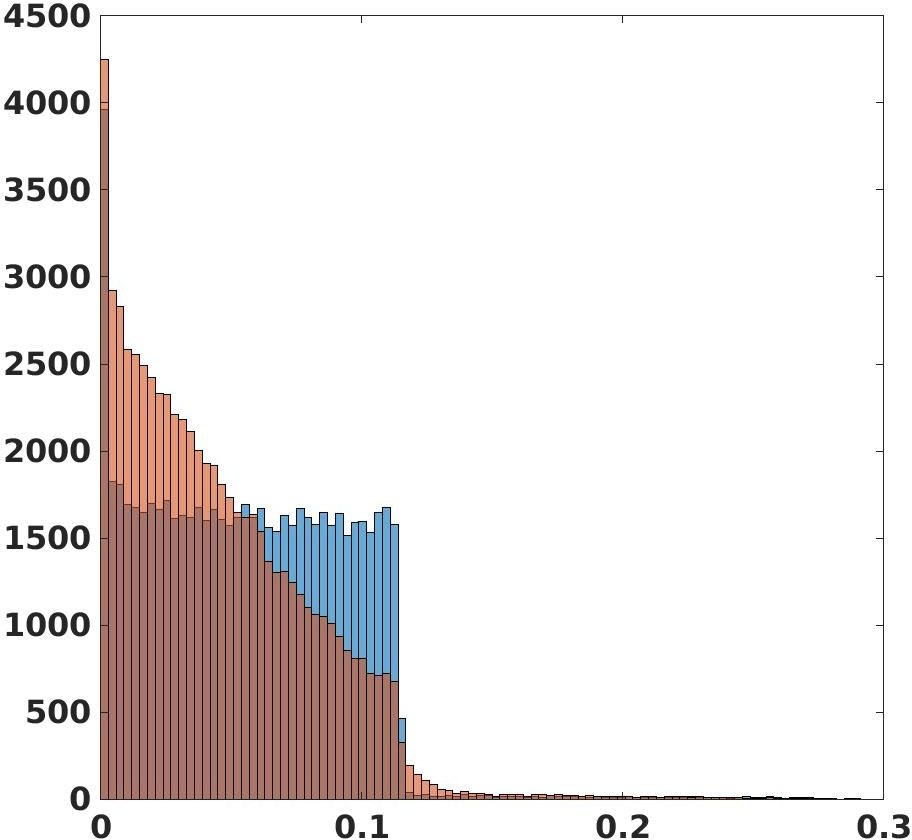} &
\hspace*{-0.375cm}
\includegraphics[width=.23\textwidth]{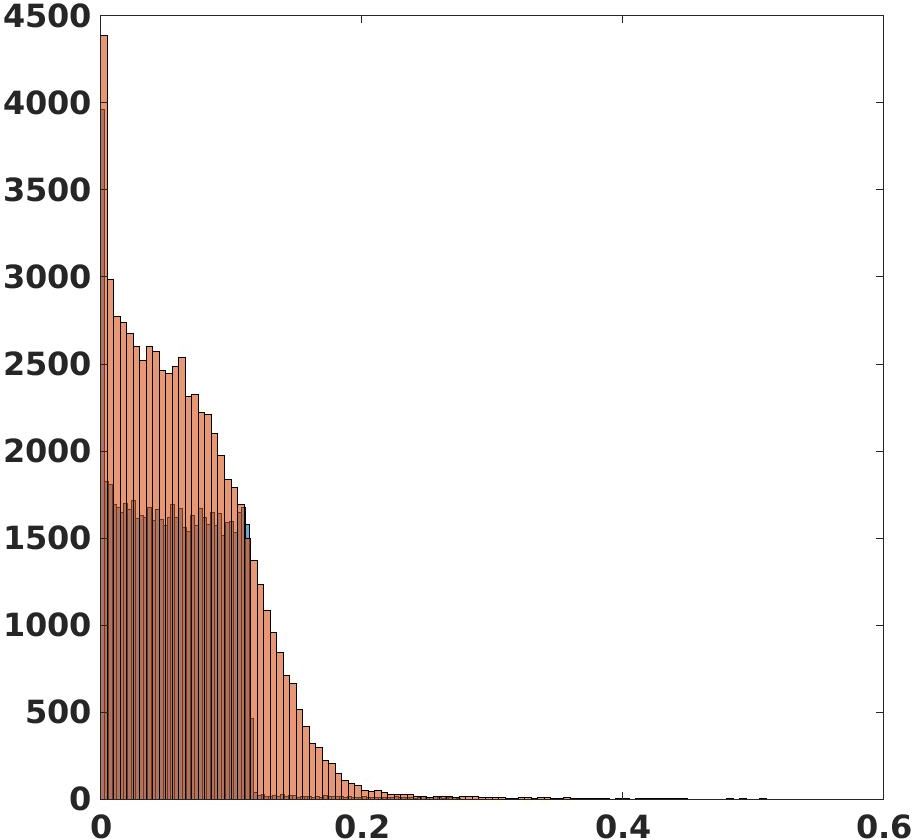} &
\hspace*{-0.375cm}
\includegraphics[width=.23\textwidth]{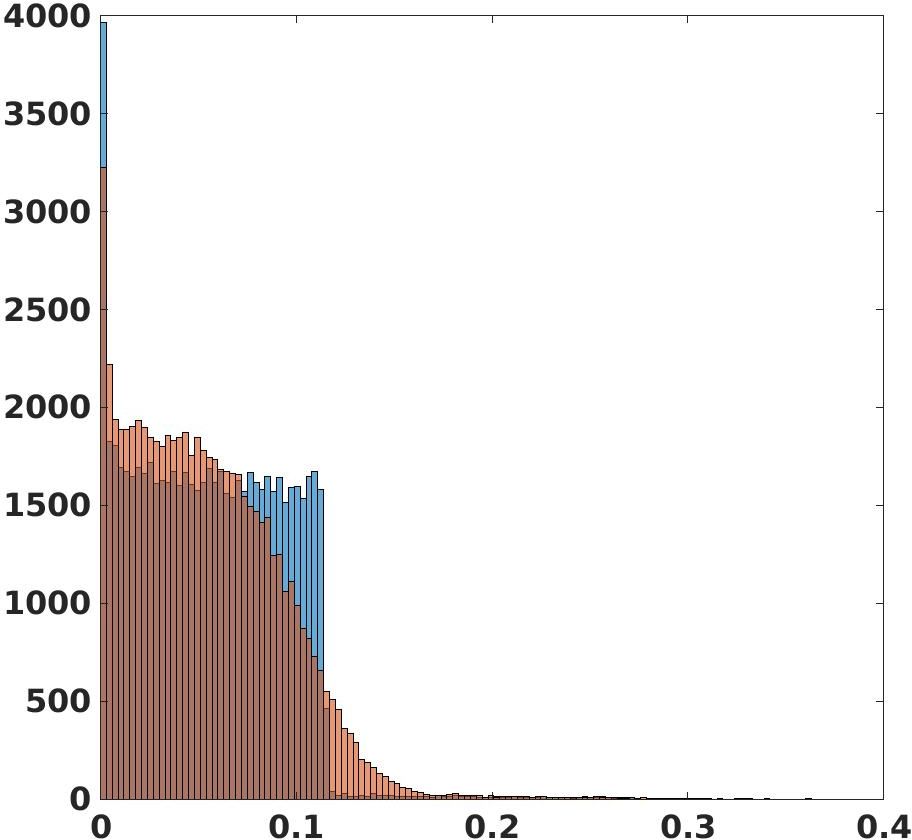} &
\hspace*{-0.375cm}
\includegraphics[width=.23\textwidth]{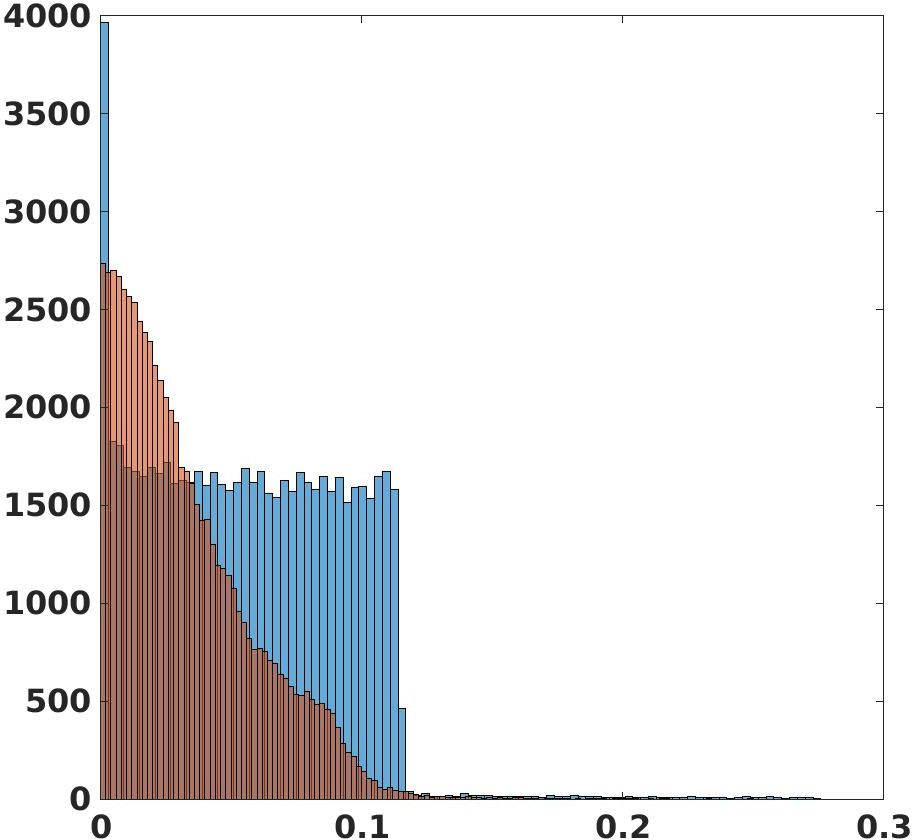}  \\
Ours & EDSR & SPM & CC \\
\end{tabular}
\caption{2X speckle noise distribution: ground-truth (blue histogram) and super-resolution method (red histogram).\label{FIG:2XSPECKLE}}
\end{figure*}

\begin{figure*}[t]
%\captionsetup{width=0.95\columnwidth}
\centering
\begin{tabular}{cc}
\includegraphics[width=.27\textwidth]{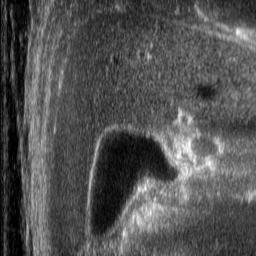} &
\includegraphics[width=.27\textwidth]{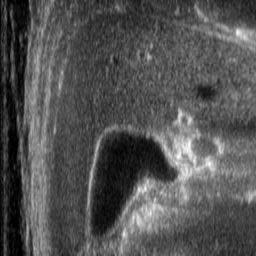} \\
US & Ours \\
\end{tabular}
\begin{tabular}{ccc}
\includegraphics[width=.27\textwidth]{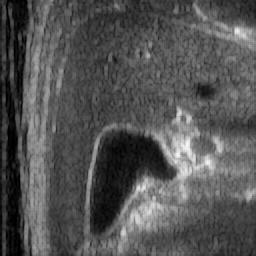} &
\includegraphics[width=.27\textwidth]{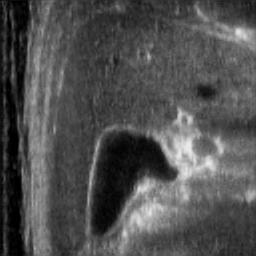} &
\includegraphics[width=.27\textwidth]{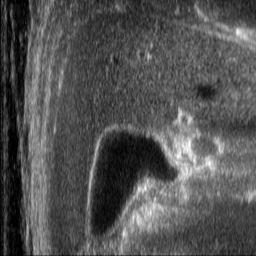}  \\
EDSR & SPM & CC
\end{tabular}
\caption{Comparison among super-resolution methods on an ultrasound image from the abdominal district, 2X up-sampling factor.\label{FIG:2XTESTUS}}
\end{figure*}

According to Tables~\ref{TAB:2XTEST} and~\ref{TAB:4XTEST}, our method has better results than state-of-the-art super-resolution methods in terms of quantitative metrics. In particular, the average MSE value of our approach with a 2X up-sampling factor is 304, while all the other methods have an average MSE value of more than 400. Also, the PSNR value of our super-resolution with the 4X up-sampling factor is 20.6, while the best result of state-of-the-art methods is 20.1, performed by cubic convolution.  Our super-resolution has better results than the other learning-based methods (i.e., EDSR and SPM) with respect to SSIM and FSIM, while CC has slightly better results. Finally, Fig.~\ref{FIG:BOXPLOTGAUSS} shows the PSNR value box plot for the 2X and 4X up-sampling factor results. Our method reduces the variability of the results of CC and outperforms EDSR and SPM.
\begin{figure*}[t]
%\captionsetup{width=0.95\columnwidth}
\centering
\begin{tabular}{ccc}
\includegraphics[width=.28\textwidth]{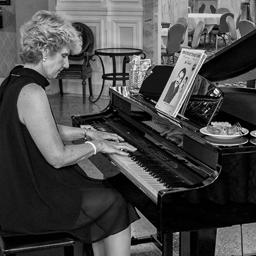} &
\includegraphics[width=.28\textwidth]{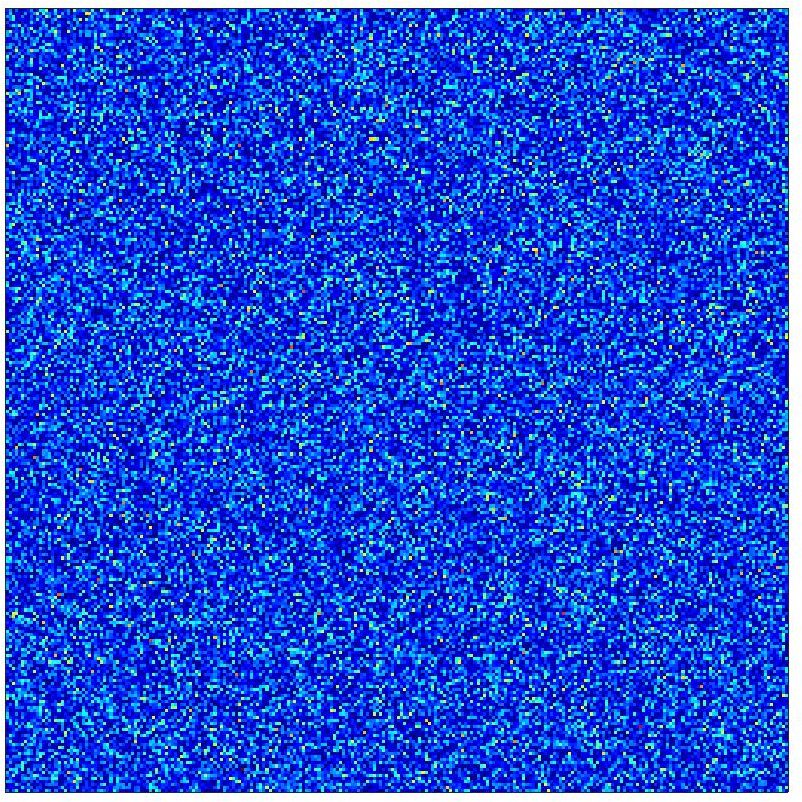} &
\includegraphics[width=.28\textwidth]{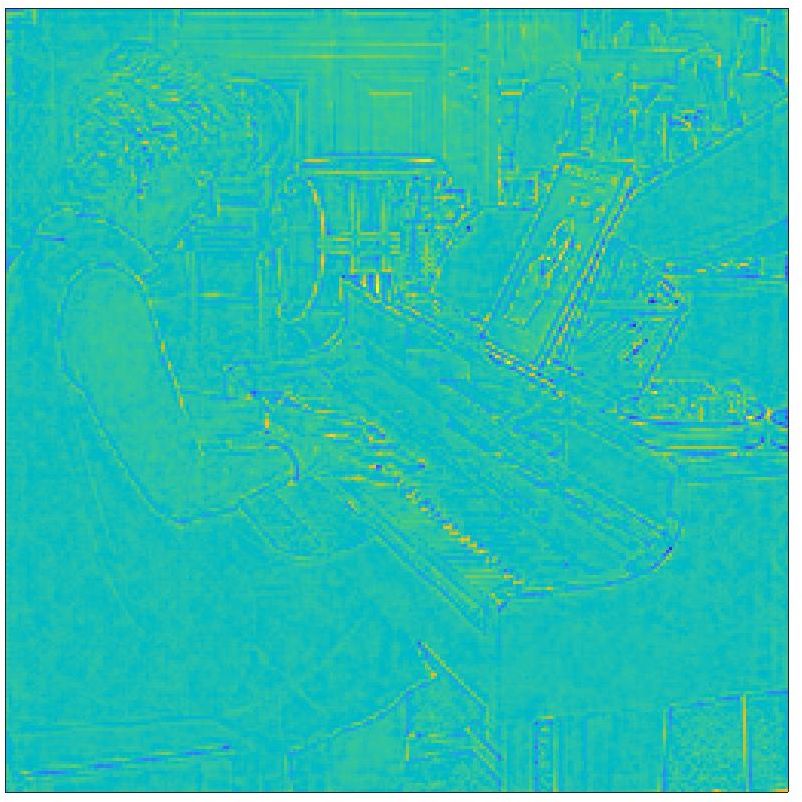} \\
Ground-truth & Noise & Ours \\
\includegraphics[width=.28\textwidth]{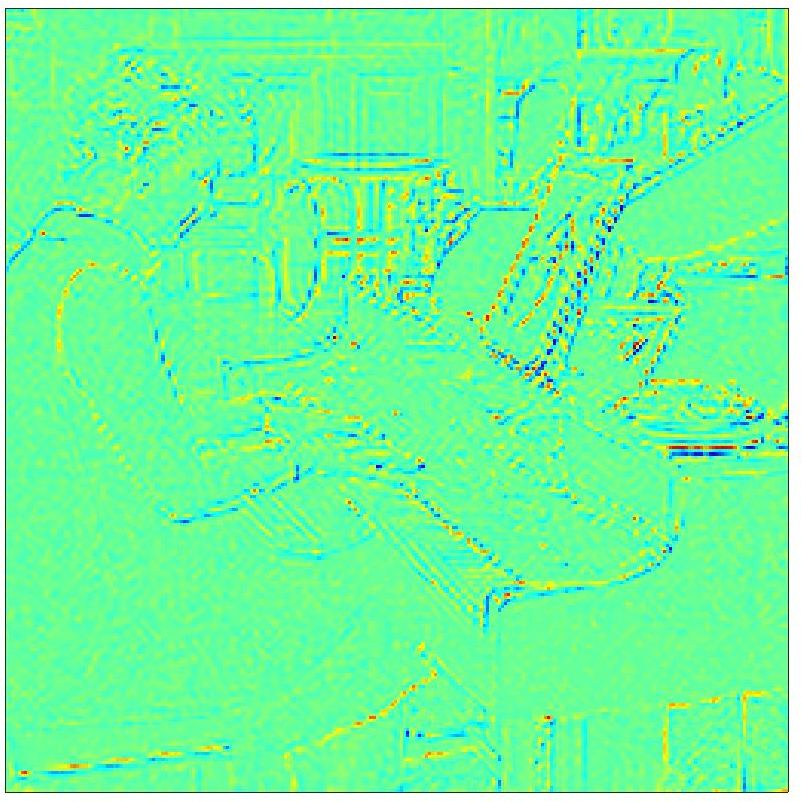} &
\includegraphics[width=.28\textwidth]{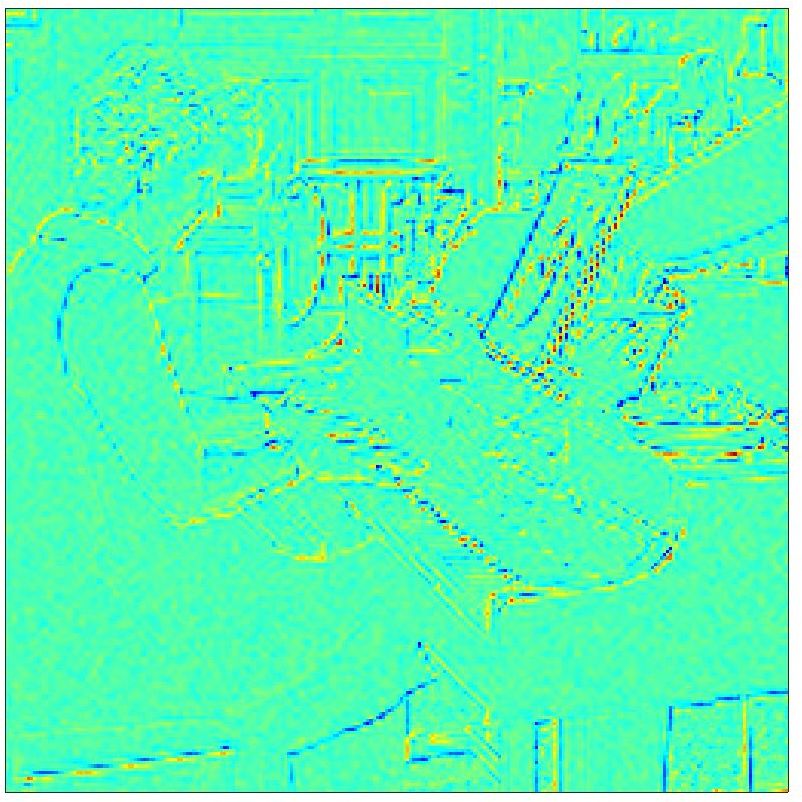} &
\includegraphics[width=.28\textwidth]{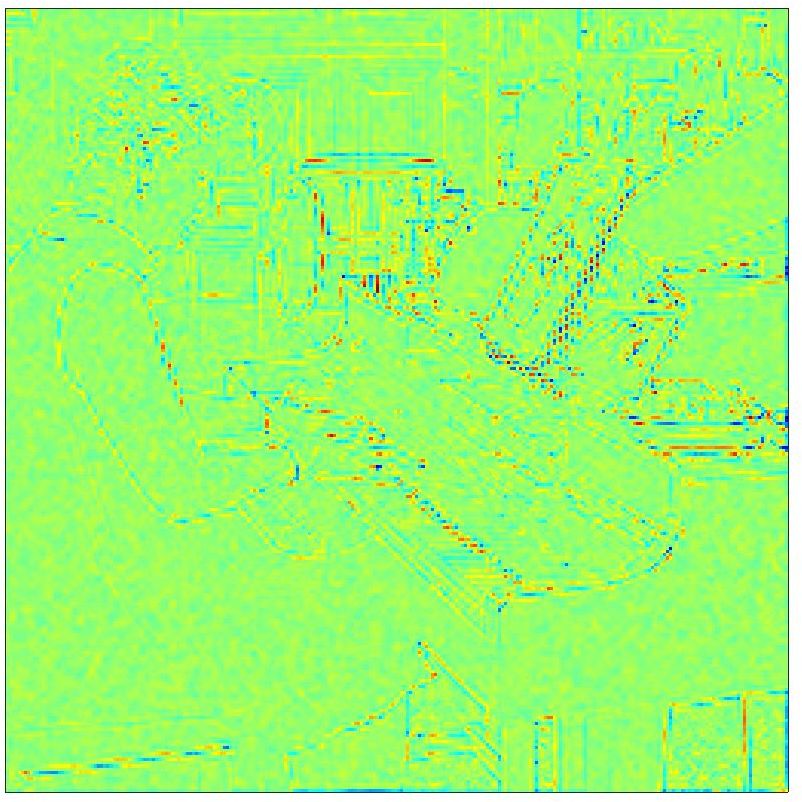}  \\
EDSR & SPM & CC
\end{tabular}
\caption{Generated noise on 2X up-sampling factor, Gaussian case.\label{FIG:2XRANDOMNOISE}}
\end{figure*}
According to the histogram (Figs.~\ref{FIG:2XGAUSS},~\ref{FIG:4XGAUSS}) of the generated noise of our versus state-of-the-art super-resolution methods, our super-resolution better preserves the distribution of the Gaussian noise in terms of mean and standard deviation, both for the 2X and 4X up-sampling factors. Fig.~\ref{FIG:2XRANDOMNOISE} shows the generated noise of the super-resolution methods with respect to the ground truth. In all the methods, the generated noise does not show good randomness properties; instead, the noise tends to adapt to the input image geometries and grey-scale values.
\begin{figure*}[t]
%\captionsetup{width=0.95\columnwidth}
\centering
\begin{tabular}{ccc}
\includegraphics[width=.27\textwidth]{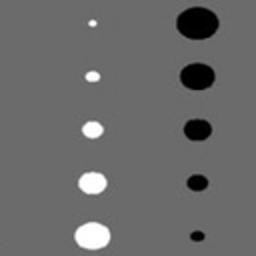} &
\includegraphics[width=.27\textwidth]{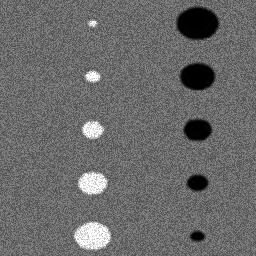} &
\includegraphics[width=.27\textwidth]{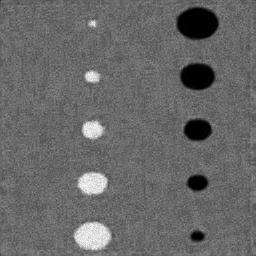} \\
Ground-truth & Noisy & Ours \\
\includegraphics[width=.27\textwidth]{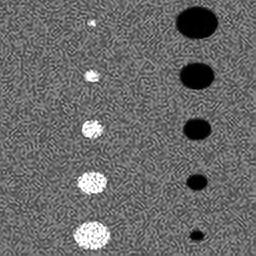} &
\includegraphics[width=.27\textwidth]{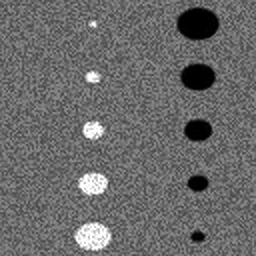} &
\includegraphics[width=.27\textwidth]{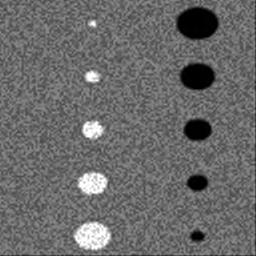}  \\
EDSR & SPM & CC
\end{tabular}
\caption{Comparison among super-resolution methods, 2X up-sampling factor, speckle noise.\label{FIG:2XTESTSPECKLE}}
\end{figure*}
\paragraph{Speckle noise}
Fig.~\ref{FIG:2XTESTSPECKLE} shows the super-resolution results with speckle noise at 2X applied on a phantom image reproducing an ultrasound scanning on cysts at various dimensions. All the methods preserve the contours of the cysts when speckle noise is used. Tables~\ref{TAB:2XTESTSPECKLE} show that our approach results better than state-of-the-art super-resolution methods in quantitative metrics. In particular, the MSE value of our method with a 2X up-sampling factor is 247, while all the other methods have a MSE value of more than 400. Our approach has comparable results with respect to other learning-based techniques (i.e., EDSR and SPM) concerning SSIM and FSIM metrics, while CC has slightly better results. CC has the worst result in preserving the noise distribution (Fig.~\ref{FIG:2XSPECKLE}), while our method has similar results to SPM. Fig.~\ref{FIG:2XTESTUS} shows the super-resolution results on an ultrasound image from the abdominal anatomical district, with 2X up-sampling. We apply our super-resolution trained with synthetic speckle noise images. Previous learning-based works generate artefacts, e.g., blurring in SPM and scattering in EDSR. CC and our method have comparable results, preserving textures and anatomical features without enhancing artefacts.

\begin{figure}[t]
%\captionsetup{width=0.95\columnwidth}
\centering
\begin{tabular}{cc}
\includegraphics[width=.45\textwidth]{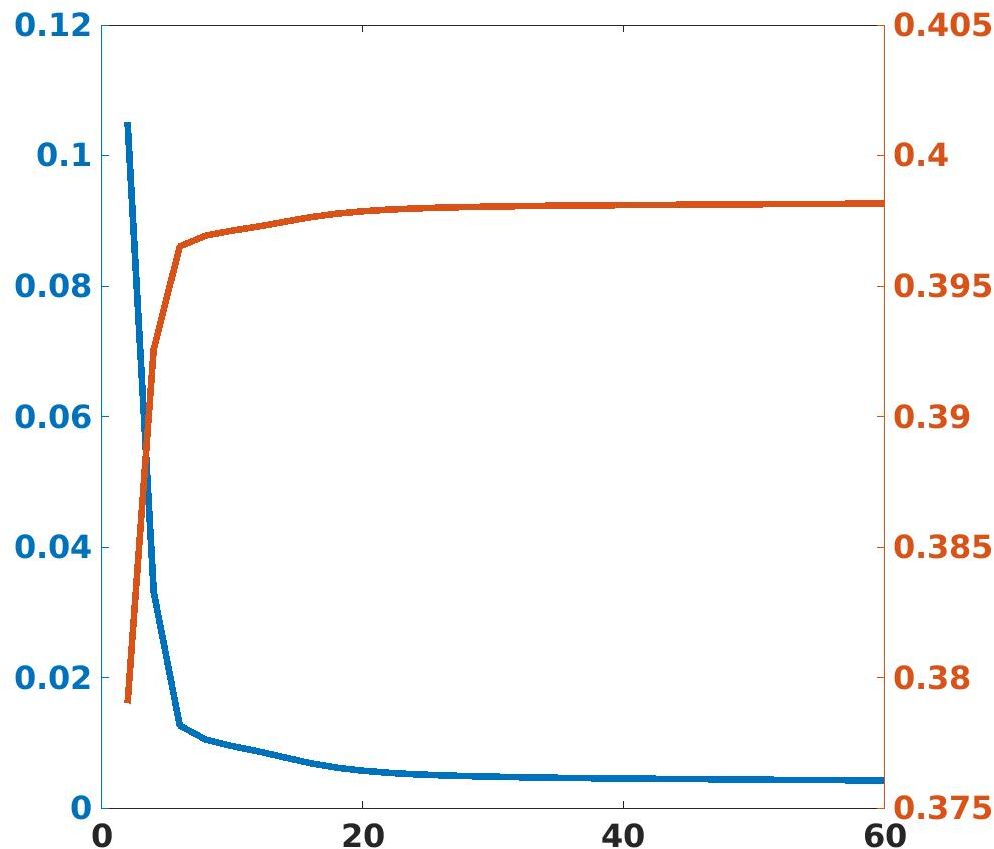} &
\includegraphics[width=.45\textwidth]{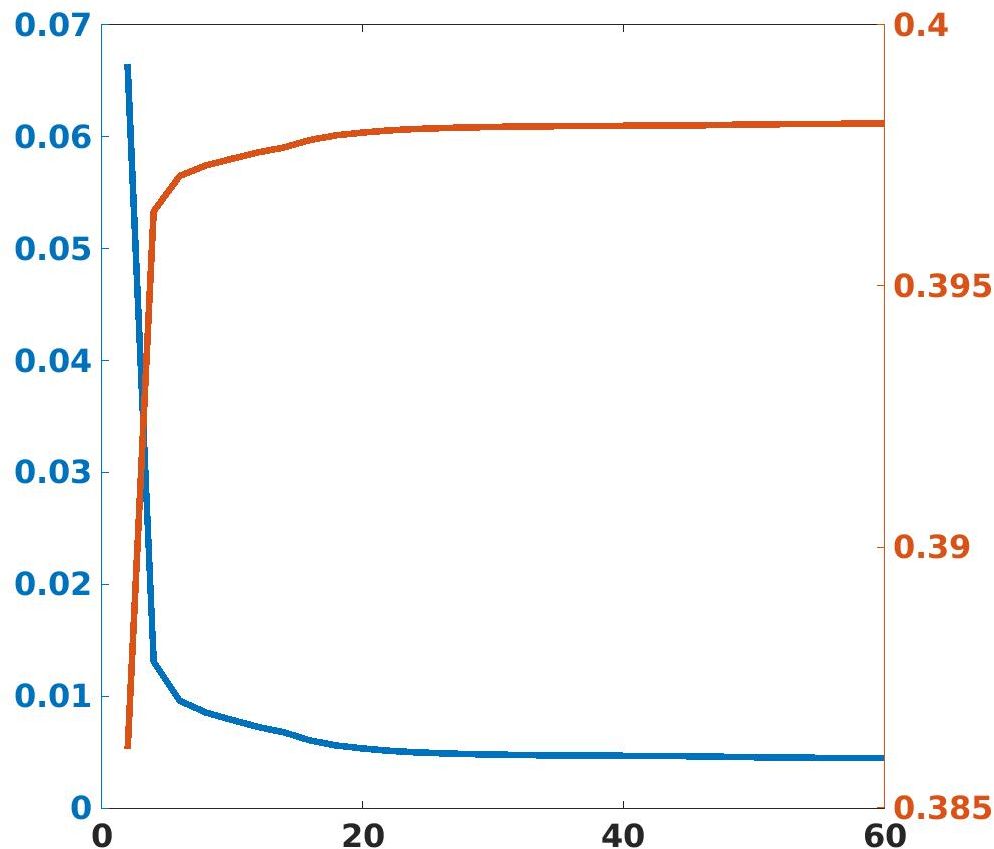} \\
Training loss &Validation loss
\end{tabular}
\caption{Training and validation loss of the Gaussian 2X network: MSE loss ($y-$axis) with respect to the number of epochs ($x-$axis). Concerning Eq.~(\ref{EQ:LOSS}), the~$y-$left axis shows the first term of the loss, while the~$y-$right axis shows the second term of the loss.   \label{FIG:LOSS}}
\end{figure}

\paragraph{Training and execution time}
Fig.~\ref{FIG:LOSS} reports the training and validation loss of the Gaussian 2X network, showing us the convergence of our model. Concerning Eq.~(\ref{EQ:LOSS}), the~$y-$left axis shows the first term of the loss (i.e.,~$\| \mathbf{P} - \mathbf{N}\|_F$), while the~$y-$right axis shows the second term of the loss (i.e., the~$\lambda$-weighted). Our training minimises the first term (i.e., the approximation norm with respect to the noisy image) and contemporary increases the second term, i.e., the probability that the generated noise complies with the input noise distribution. We mention that the execution time of the training is around 30 seconds per epoch on the Cineca Marconi100 cluster, at the 26th position in the ``\emph{top500}'' list~\cite{urlcineca}. The cluster uses 980 nodes, each with IBM Power9 AC922 at 3.1GHz 32 cores and 4 NVIDIA Volta V100 GPUs per node, with the GPU interconnection NVlink 2.0 at 16GB and 256GB of RAM each node. The prediction time is lower than 1 second on a standard workstation with 2 Intel i9-9900KF CPUs (3.60GHz), 32GB RAM, and Tensorflow 2.7. 
\begin{table}[t]
%\captionsetup{width=0.95\columnwidth}
\centering
\caption{Concerning the 2X up-sampling factor results on speckle noise images, we report the metrics computed between target and super-resolution methods as the average value on the test data set. The best results are in bold.\label{TAB:2XTESTSPECKLE}}
\begin{tabular}{c|cccc}
Methods & CC & EDSR & SPM & OUR \\ \hline
MSE & 408.81 & 600.21 & 488.39  &~$\mathbf{247.67}$\\
NRMSE &0.1832 & 0.220 & 0.199 &~$\mathbf{0.121}$ \\
NCC & 0.812 & 0.744& 0.780&~$\mathbf{0.853}$ \\
PSNR &22.01 & 20.35 &21.24 &$\mathbf{24.19}$ \\
SSIM &~$\mathbf{0.365}$ &0.301& 0.318 &0.312 \\
FSIM &~$\mathbf{0.875}$ & 0.819 & 0.854 &0.818 \\
UIQ & 0.988 & 0.982 & 0.985 &~$\mathbf{0.990}$
\end{tabular}
\end{table}
\section{Conclusion and future work\label{SEC:CONCLUSION}}
We have presented our novel learning-based method for 2D noisy images super-resolution. Our approach reconstructs the high-resolution image while preserving noise-related geometries and features. We have tested our super-resolution with different up-sampling factors (e.g., 2X and 4X) and noise types (e.g., Gaussian, speckle), comparing quantitative results with state-of-the-art methods. Our method outperforms learning-based methods and has comparable results with standard methods. In future works, we plan to extend our super-resolution method to applicative contexts (e.g., biomedical) where the noise distribution is unknown.

{\small{\paragraph{\textbf{Acknowledgements}} 
This work has been partially supported by the European Commission, NextGenerationEU, Missione 4 Componente 2, ``\emph{Dalla ricerca all’impresa}'', Innovation Ecosystem RAISE ``\emph{Robotics and AI for Socio-economic Empowerment}'', ECS00000035. Tests on the CINECA Cluster are supported by the ISCRA-C project US-SAMP, HP10CXLQ1S.}}
%
% {\small{\paragraph{\textbf{Conflict of interest}}
 %The authors declare no competing interests.}} 
%--------------------------------
\bibliographystyle{alpha}
\bibliography{refs}

\newcommand{\etalchar}[1]{$^{#1}$}
\begin{thebibliography}{WYW{\etalchar{+}}19}

\bibitem[CNP22]{PMID:35672630}
Simone Cammarasana, Paolo Nicolardi, and Giuseppe Patanè.
\newblock Real-time denoising of ultrasound images based on deep learning.
\newblock {\em Medical \& Biological Engineering \& Computing},
  60(8):2229—2244, August 2022.

\bibitem[CNP23]{cammarasana2023super}
Simone Cammarasana, Paolo Nicolardi, and Giuseppe Patane.
\newblock Super-resolution of 2{D} ultrasound images and videos.
\newblock {\em Medical \& Biological Engineering \& Computing}, 05 2023.

\bibitem[DLHT14]{dong2014learning}
Chao Dong, Chen~Change Loy, Kaiming He, and Xiaoou Tang.
\newblock Learning a deep convolutional network for image super-resolution.
\newblock In {\em European Conference on Computer Vision}, pages 184--199.
  Springer, 2014.

\bibitem[DLHT15]{dong2015image}
Chao Dong, Chen~Change Loy, Kaiming He, and Xiaoou Tang.
\newblock Image super-resolution using deep convolutional networks.
\newblock {\em IEEE Transactions on Pattern Analysis and Machine Intelligence},
  38(2):295--307, 2015.

\bibitem[GJB03]{gribbon2003real}
KT~Gribbon, CT~Johnston, and Donald~G Bailey.
\newblock A real-time {FPGA} implementation of a barrel distortion correction
  algorithm with bilinear interpolation.
\newblock In {\em Image and Vision Computing New Zealand}, pages 408--413,
  2003.

\bibitem[HLH21]{han2021Deep}
Sujy Han, Tae~Bok Lee, and Yong~Seok Heo.
\newblock Deep image prior for super resolution of noisy image.
\newblock {\em Electronics}, 10(16):2014, Aug 2021.

\bibitem[HSA15]{huang2015single}
Jia-Bin Huang, Abhishek Singh, and Narendra Ahuja.
\newblock Single image super-resolution from transformed self-exemplars.
\newblock In {\em Proceedings of the IEEE Conference on Computer Vision and
  Pattern Recognition}, pages 5197--5206, 2015.

\bibitem[KAJ{\etalchar{+}}20]{khaledyan2020low}
Donya Khaledyan, Abdolah Amirany, Kian Jafari, Mohammad~Hossein Moaiyeri,
  Abolfazl~Zargari Khuzani, and Najmeh Mashhadi.
\newblock Low-cost implementation of bilinear and bicubic image interpolation
  for real-time image super-resolution.
\newblock In {\em Global Humanitarian Technology Conference}, pages 1--5. IEEE,
  2020.

\bibitem[Key81]{keys1981cubic}
Robert Keys.
\newblock Cubic convolution interpolation for digital image processing.
\newblock {\em Transactions on Acoustics, Speech, and Signal Processing},
  29(6):1153--1160, 1981.

\bibitem[Kov99]{kovesi1999image}
Peter Kovesi.
\newblock Image features from phase congruency.
\newblock {\em Videre: Journal of Computer Vision Research}, 1(3):1--26, 1999.

\bibitem[LAPB14]{lapini2014comparison}
Alessandro Lapini, Fabrizio Argenti, Alessandro Piva, and Luca Bencini.
\newblock Comparison of super-resolution methods for quality enhancement of
  digital biomedical images.
\newblock In {\em 2014 8th International symposium on medical information and
  communication technology (ISMICT)}, pages 1--5. IEEE, 2014.

\bibitem[LSK{\etalchar{+}}17]{lim2017enhanced}
Bee Lim, Sanghyun Son, Heewon Kim, Seungjun Nah, and Kyoung Mu~Lee.
\newblock Enhanced deep residual networks for single image super-resolution.
\newblock In {\em Proc. of the IEEE Conf. on Computer Vision and Pattern
  Recognition Workshops}, pages 136--144, 2017.

\bibitem[LTH{\etalchar{+}}17]{ledig2017photo}
Christian Ledig, Lucas Theis, Ferenc Huszar, Jose Caballero, Andrew Cunningham,
  Alejandro Acosta, Andrew Aitken, Alykhan Tejani, Johannes Totz, Zehan Wang,
  and Wenzhe Shi.
\newblock Photo-realistic single image super-resolution using a generative
  adversarial network.
\newblock In {\em Proceedings of the IEEE Conference on Computer Vision and
  Pattern Recognition}, pages 4681--4690, 2017.

\bibitem[MMP{\etalchar{+}}14]{mahale2014hardware}
Gopinath Mahale, Hamsika Mahale, Rajesh~Babu Parimi, SK~Nandy, and
  S~Bhattacharya.
\newblock Hardware architecture of bi-cubic convolution interpolation for
  real-time image scaling.
\newblock In {\em International Conference on Field-Programmable Technology},
  pages 264--267. IEEE, 2014.

\bibitem[MNV{\etalchar{+}}22]{mushtaq2022super}
Zaid Mushtaq, Shoaib Nasti, Chaman Verma, Maria Raboaca, Neerendra Kumar, and
  Samiah Nasti.
\newblock Super resolution for noisy images using convolutional neural
  networks.
\newblock {\em Mathematics}, 10(5):777, Feb 2022.

\bibitem[PE14]{peleg2014statistical}
Tomer Peleg and Michael Elad.
\newblock A statistical prediction model based on sparse representations for
  single image super-resolution.
\newblock {\em Transactions on Image Processing}, 23(6):2569--2582, 2014.

\bibitem[PK05]{puschmann2005super}
Klaus~G Puschmann and Franz Kneer.
\newblock On super-resolution in astronomical imaging.
\newblock {\em Astronomy \& Astrophysics}, 436(1):373--378, 2005.

\bibitem[QCJY22]{qin2022progressive}
Jiayi Qin, Lihui Chen, Seunggil Jeon, and Xiaomin Yang.
\newblock Progressive interaction-learning network for lightweight single-image
  super-resolution in industrial applications.
\newblock {\em IEEE Transactions on Industrial Informatics}, 19(2):2183--2191,
  2022.

\bibitem[RDS{\etalchar{+}}15]{russakovsky2015imagenet}
Olga Russakovsky, Jia Deng, Hao Su, Jonathan Krause, Sanjeev Satheesh, Sean Ma,
  Zhiheng Huang, Andrej Karpathy, Aditya Khosla, Michael Bernstein,
  Alexander~C. Berg, and Li~Fei-Fei.
\newblock Imagenet large scale visual recognition challenge.
\newblock {\em International Journal of Computer Vision}, 115:211--252, 2015.

\bibitem[RR20]{rakotonirina2020esrgan+}
Nathana{\"e}l~Carraz Rakotonirina and Andry Rasoanaivo.
\newblock Esrgan+: Further improving enhanced super-resolution generative
  adversarial network.
\newblock In {\em ICASSP 2020-2020 International Conference on Acoustics,
  Speech and Signal Processing}, pages 3637--3641. IEEE, 2020.

\bibitem[SPP15]{salvador2015naive}
Jordi Salvador and Eduardo Perez-Pellitero.
\newblock Naive {B}ayes super-resolution forest.
\newblock In {\em Proceedings of the IEEE International Conference on Computer
  Vision}, pages 325--333, 2015.

\bibitem[SS20]{singh2020survey}
Amanjot Singh and Jagroop Singh.
\newblock Survey on single image based super-resolution—implementation
  challenges and solutions.
\newblock {\em Multimedia Tools and Applications}, 79:1641--1672, 2020.

\bibitem[TDSVG13]{timofte2013anchored}
Radu Timofte, Vincent De~Smet, and Luc Van~Gool.
\newblock Anchored neighborhood regression for fast example-based
  super-resolution.
\newblock In {\em Proceedings of the IEEE International Conference on Computer
  Vision}, pages 1920--1927, 2013.

\bibitem[TDSVG14]{timofte2014a+}
Radu Timofte, Vincent De~Smet, and Luc Van~Gool.
\newblock A+: adjusted anchored neighborhood regression for fast
  super-resolution.
\newblock In {\em Asian Conference on Computer Vision}, pages 111--126.
  Springer, 2014.

\bibitem[url]{urlcineca}
{Cineca Marconi100}.
\newblock \url{https://www.top500.org/system/179845/}.
\newblock Accessed: 2023-08-01.

\bibitem[VCSR21]{villar2021deep}
Angel Villar-Corrales, Franziska Schirrmacher, and Christian Riess.
\newblock Deep learning architectural designs for super-resolution of noisy
  images.
\newblock pages 1635--1639, 06 2021.

\bibitem[WB02]{wang2002universal}
Zhou Wang and Alan~C Bovik.
\newblock A universal image quality index.
\newblock {\em Signal Processing Letters}, 9(3):81--84, 2002.

\bibitem[WCH20]{wang2020deep}
Zhihao Wang, Jian Chen, and Steven~CH Hoi.
\newblock Deep learning for image super-resolution: a survey.
\newblock {\em IEEE Transactions on Pattern Analysis and Machine Intelligence},
  43(10):3365--3387, 2020.

\bibitem[WYW{\etalchar{+}}19]{10.1007/978-3-030-11021-5_5}
Xintao Wang, Ke~Yu, Shixiang Wu, Jinjin Gu, Yihao Liu, Chao Dong, Yu~Qiao, and
  Chen~Change Loy.
\newblock Esrgan: Enhanced super-resolution generative adversarial networks.
\newblock In Laura Leal-Taix{\'e} and Stefan Roth, editors, {\em Computer
  Vision -- ECCV 2018 Workshops}, pages 63--79, Cham, 2019. Springer
  International Publishing.

\bibitem[YFH20]{yu2020wide}
Jiahui Yu, Yuchen Fan, and Thomas Huang.
\newblock Wide activation for efficient image and video super-resolution.
\newblock In {\em 30th British Machine Vision Conference, BMVC 2019}, 2020.

\bibitem[YWHM10]{yang2010image}
Jianchao Yang, John Wright, Thomas~S Huang, and Yi~Ma.
\newblock Image super-resolution via sparse representation.
\newblock {\em IEEE Transactions on Image Processing}, 19(11):2861--2873, 2010.

\bibitem[ZFB{\etalchar{+}}18]{zhang2018single}
Yunfeng Zhang, Qinglan Fan, Fangxun Bao, Yifang Liu, and Caiming Zhang.
\newblock Single-image super-resolution based on rational fractal
  interpolation.
\newblock {\em IEEE Transactions on Image Processing}, 27(8):3782--3797, 2018.

\bibitem[ZZMZ11]{zhang2011fsim}
Lin Zhang, Lei Zhang, Xuanqin Mou, and David Zhang.
\newblock Fsim: A feature similarity index for image quality assessment.
\newblock {\em Transactions on Image Processing}, 20(8):2378--2386, 2011.

\end{thebibliography}
%
%----BIOGRAPHY---
\begin{description}
\item[\textbf{Simone Cammarasana}] is researcher at CNR-IMATI. He obtained a PhD in Computer Science at the University of Genova-DIBRIS, a post-lauream Master in Scientific Computing at the University of Sapienza-Roma, and a Master's degree in Engineering at the University of Pisa. His research interests include signals analysis, optimisation problems, and medical images.

\item[\textbf{Giuseppe Patan\'e}] is senior researcher at CNR-IMATI. Since 2001, his research is mainly focused on Computer Graphics and Shape Modelling. He is the author of scientific publications in international journals and conference proceedings, and a tutor of PhD and Post.Doc students. He is responsible for R$\&$D activities in national and European projects.
\end{description}
\end{document}